\documentclass[a4paper,fleqn]{cas-sc}

\usepackage[authoryear]{natbib}
\graphicspath{{./}{figures/}}

\def\tsc#1{\csdef{#1}{\textsc{\lowercase{#1}}\xspace}}
\tsc{WGM}
\tsc{QE}

\begin{document}
\let\WriteBookmarks\relax
\def\floatpagepagefraction{1}
\def\textpagefraction{.001}

\shorttitle{A critical assessment of the ``energy limited'' argument}    

\shortauthors{D. Modirrousta-Galian and J. Korenaga}  

\title [mode = title]{Revising core powered mass loss: A critical assessment of the ``energy limited'' argument}  



%

\author[1,2,3]{Darius Modirrousta-Galian}

\cormark[1]

\ead{modirrousta-galian@sjtu.edu.cn}



\affiliation[1]{organization={Tsung‐Dao Lee Institute, Shanghai Jiao Tong University},
            city={Shanghai},
            country={China}}

\affiliation[2]{organization={School of Physics and Astronomy, Shanghai Jiao Tong University},
            city={Shanghai},
            country={China}}

\affiliation[3]{organization={Yale University, Department of Earth and Planetary Sciences},
            city={New Haven},
            state={CT},
            country={USA}}

\author[3]{Jun Korenaga}

\ead{}




\cortext[1]{Darius Modirrousta-Galian}

\fntext[1]{}


\begin{abstract}
The extreme conditions in the early stages of planetary evolution are thought to shape its subsequent development. High internal temperatures from giant impacts can provide sufficient energy to drive extreme volatile loss, with hydrogen being most readily lost. However, the conditions required for maintaining a primordial atmosphere over geological timescales remain enigmatic. This paper revisits the core powered mass loss model for hydrogen removal from planetary atmospheres. One popular approach is to combine mass continuity at the sonic point with an energy-based constraint. We demonstrate that the so-called ``energy limited'' component of this model is unnecessary because atmospheric loss following giant impacts is governed solely by conditions at the sonic point. By simulating a broad range of synthetic exoplanets, varying in planetary mass, atmospheric mass fraction, and temperature, we find that the ``energy limited'' model can underestimate the mass loss rates by up to eight orders of magnitude. Our findings suggest that, for sufficiently hot post-impact surface conditions, hydrogen rich atmospheres can be removed on dynamical timescales that are far shorter than one million years.
\end{abstract}


\begin{highlights}
\item The core powered mass loss rate is not limited by planetary internal luminosity through the radiative-convective boundary. 
\item The core powered mass loss rate depends solely on the conditions at the sonic point.
\item The revised core powered mass loss model yields significantly higher mass loss rates by up to eight orders of magnitude. 
\item The revised mass loss rate is well approximated by $\dot{M}{\approx}1.26{\times}10^{15}(R_{\rm s}/R_\oplus)^{2}\exp{\left[{-}19R_{\rm s}/(27R_\oplus)\right]}\,{\rm kg/s}$, where $R_{\rm s}$ is the sonic point.
\end{highlights}


\maketitle

\section{Introduction} 
\label{sec:intro}

From observations, it is inferred that many super-Earths and sub-Neptunes \citep[i.e., planets with radii ${\lesssim}3.5~R_{\oplus}$;][]{Fulton2017} lose most or all of their atmospheric hydrogen early in their evolutions \citep{Mazeh2016,Fulton2017,Fulton2018}. Hydrogen can be lost readily from planetary atmospheres because its low mass allows for it to become gravitationally unbound if it is sufficiently heated \citep{Jeans1925,Chamberlain1962,Opik1963}. The manner by which it is lost is less clear; the two most widely cited mechanisms are external heating from X-ray and ultraviolet (XUV) irradiation \citep{Watson1981,Zahnle1986,Chassefiere1996,Erkaev2007,Murray2009,Owen2019,Modirrousta2024b} and internal heating from giant impacts \citep{Walker1986,Ahrens1993,Ginzburg2018,Biersteker2019,Modirrousta2023}. Both mechanisms have been shown to potentially explain the bimodal distribution of exoplanet radii, i.e., the two maxima at $1.3~R_{\oplus}$ and $2.4~R_{\oplus}$, and the minimum at $1.75~R_{\oplus}$ \citep{Owen2017,Ginzburg2018}.

For both mechanisms, mass loss has usually been modeled using ``energy limited'' arguments. These arguments generally state that energy is required to do work against gravity in transporting atmospheric gas to a distance where it can overcome the gravitational pull of the planet. The specific energy source depends on the mass loss mechanism under consideration. For XUV-induced photoevaporation, the equation is \citep{Erkaev2007,Owen2013,Owen2017,Kubyshkina2018(2),Kubyshkina2021},
\begin{equation}
    \left.\frac{{\rm d}M_{\rm atm}}{{\rm d}t}\right|_{\rm XUV} = \frac{\pi R_{\rm p} R_{\rm XUV}^{2}\varepsilon F_{\rm XUV}}{GM_{\rm p}K},
\label{eq:XUV_EL}
\end{equation}
where $R_{\rm p}$ is the planetary photospheric radius (i.e., where the average optical depth $\tau{\approx}2/3$), $R_{\rm XUV}$ is the XUV-absorption radius (i.e., where the XUV optical depth $\tau_{\rm XUV}{\approx}2/3$), $\varepsilon$ is the thermodynamic heating efficiency \citep[see section~6;][]{Modirrousta2023}, $F_{\rm XUV}$ is the incoming stellar XUV flux, $G$ is the gravitational constant, $M_{\rm p}$ is the total planetary mass, and $K$ is a factor very close to unity that accounts for the effects of the Hill sphere/Roche lobe radius \citep{Erkaev2007}. The argument is that incoming XUV energy is converted into work done against gravity in transporting gas from the planetary photosphere to the Hill sphere where it is lost. The photosphere is selected as the lower boundary because it marks the bottom of the thermosphere, where gas heats up sharply because of conductive heat transfer from above \citep{Bates1951,Bates1959,Bauer1971,Gross1972,Horedt1982}. Equation~\ref{eq:XUV_EL} relies on several key assumptions that have been recognized by subsequent studies: (1) \citet{Krenn2021} challenged both the quasistatic mass loss assumption and the assumption that the XUV-absorption radius is always below the sonic point, (2) \citet{Kubyshkina2018(1)} showed that planetary equilibrium temperature cannot be ignored, (3) \citet{Salz2016} demonstrated that a sonic point does not always exist, and (4) \citet{Modirrousta2024b} showed that this framework overlooks viscous effects between hydrogen and heavier gases, overestimating mass loss rates by several orders of magnitude. This effect is small when heavier species are only trace constituents, but it becomes increasingly important as hydrogen is preferentially lost and the relative abundance of heavier species increases, providing an intrinsic negative feedback on mass loss. The same compositional shift may also increase the abundance of efficient infrared radiators such as CO$_2$, lowering thermospheric temperatures and further suppressing escape rates \citep[e.g.,][]{Johnstone2021}.

For mass loss from internal heating caused by giant impacts, the ``energy limited'' equation is \citep{Ginzburg2018,Gupta2018,Biersteker2019,Gupta2020},
\begin{equation}
    \left.\frac{{\rm d}M_{\rm atm}}{{\rm d}t}\right|_{\rm GI{,}1} = \frac{\epsilon L_{\rm int}R_{\rm rcb}}{GM_{\rm p}},
\label{eq:GI_EL1}
\end{equation}
where $L_{\rm int}$ is the internal luminosity of the planet, $\epsilon$ is the heating efficiency, and $R_{\rm rcb}$ is the radius of the radiative-convective boundary (see Figure~\ref{fig:schematic}). This boundary is chosen because gas above is stably stratified, and energy is therefore required to move mass radially outward. It is suggested that internal energy is used to lift mass from the radiative-convective boundary to infinity (or the Hill sphere if parameter $K$ were incorporated). The subscript 1 is included because there are two equations involved in the standard core powered mass loss framework, which we discuss later in the manuscript. Despite its similarity to Equation~\ref{eq:XUV_EL}, the validity of Equation~\ref{eq:GI_EL1} remains mostly unexplored. Here, we provide this missing evaluation of the ``energy limited'' argument in the standard core powered mass loss model.

In this context, it is essential to recognize that the core powered mass loss framework in which Equation~\ref{eq:GI_EL1} is applied is physically distinct from the regime of XUV-induced photoevaporation. Whereas XUV-induced photoevaporation and core powered mass loss have been explored in parallel in the literature \citep[e.g.,][]{Kubyshkina2021,Matsumoto2021,Zhong2024}, it is often not appreciated that these mechanisms operate in physically distinct and mutually exclusive atmospheric regimes \citep{Modirrousta2023}. Core powered mass loss occurs when the entire atmosphere below the sonic point is optically thick to visible and thermal photons, so that mass loss is powered by internal energy or a mixture of internal and thermal stellar energy. In this state, stellar X-ray and ultraviolet photons are absorbed above the sonic point, where they cannot influence the deeper atmosphere. Because information such as heat and pressure perturbations can propagate only at the local sound speed, a transonic outflow prevents these perturbations from traveling downward against the flow. Conversely, X-ray and ultraviolet driven escape can only operate when a thermosphere is present, that is, when the upper atmosphere is optically thin to visible and thermal photons but remains optically thick to high-energy irradiation. The two frameworks are not additive and cannot be applied to the same planet at the same time \citep[][section 5.2.]{Modirrousta2023}; instead, X-ray and ultraviolet induced photoevaporation may act during intervals when a thermosphere persists between successive giant impacts.

Previous work has demonstrated that the ``energy limited'' approach does not provide an accurate description of XUV-induced photoevaporation for planets of low mass, high equilibrium temperature, or subject to intense XUV fluxes \citep{Lammer2016,Kubyshkina2018(1),Kubyshkina2018(2),Kubyshkina2021}. Here, we build on this approach by examining whether the ``energy limited'' argument remains valid in the regime where the entire atmosphere is optically thick and mass loss is powered by internal energy, i.e., in the core powered framework. We show that the mass loss rate is likewise controlled by the hydrodynamic structure of the flow, which is defined by the sonic point conditions, rather than by an energy limit set by the interior cooling rate. This outcome is consistent with standard hydrodynamic analyses for strongly irradiated planets heated from above \citep[e.g.,][]{Lammer2016}. However, this connection is often overlooked in treatments of core powered mass loss, where heating is from below and ``energy limited'' arguments are frequently adopted instead \citep[e.g.,][their Equation~12]{Owen2016}. Recognizing this parallel allows us to reinterpret the core powered regime within the same hydrodynamic framework, highlighting that the escape rate is set by sonic point conditions rather than by the planet’s luminosity.

This paper is structured as follows. Section~\ref{sec:review} reviews the most widely cited version of the core powered mass loss model. Section~\ref{sec:assumptions} examines the underlying assumptions of the ``energy limited'' argument. Section~\ref{sec:outlook} revises the core powered mass loss framework and compares it with the standard approach. Section~\ref{sec:approximation} provides an efficient approximation for the core powered mass loss rate, suitable for population-level analyses. Section~\ref{sec:conclusions} summarizes our findings.

\section{Review of the core powered mass loss model}
\label{sec:review}

The core powered mass loss model was introduced in the late 2010s \citep{Ginzburg2016,Ginzburg2018,Gupta2018,Biersteker2019,Gupta2020}. It builds on canonical planetary formation theory, which suggests that planets the size of Earth or larger experience multiple giant impacts during their final phase of accretion. During this phase, massive collisions generate extremely high surface temperatures \citep[e.g., $T{>}10,000~{\rm K}$;][]{Benz1990,Cameron1997,Canup2008,Nakajima2015,Lock2018}. We define a giant impact as a collision in which a planetary body is struck by another body of comparable or smaller mass during late-stage accretion \citep[][see their section~2.4.4.2]{Chambers2014}. Such events occur over a broad mass range and typically at velocities comparable to the mutual escape speed \citep{Agnor1999,Kokubo2007,Chambers2013,Suli2021,Matsumoto2021}. In this paper we do not model the physics or statistics of giant impacts themselves, with our focus being only on the fluid dynamics of atmospheric escape in the context of core powered mass loss.

If the primary body hosts a primordial hydrogen-rich atmosphere before the collision, this atmosphere may be lost if it is no longer gravitationally bound after the collision. For very large planets like gas giants, the gas is strongly bound and it is not lost through this mechanism. The core powered mass loss model is thus considered only for super-Earths and sub-Neptunes, which usually have masses below $10~M_{\oplus}$. For simplicity, it is assumed that the atmospheric mass is negligible compared to that of the condensed section of the planet (the planetary nucleus henceforth) so that $M_{\rm p}{=}M_{\rm n}{+}M_{\rm atm}{\approx}M_{\rm n}$.

\begin{figure}
    \centering
    \includegraphics[width=0.6\linewidth]{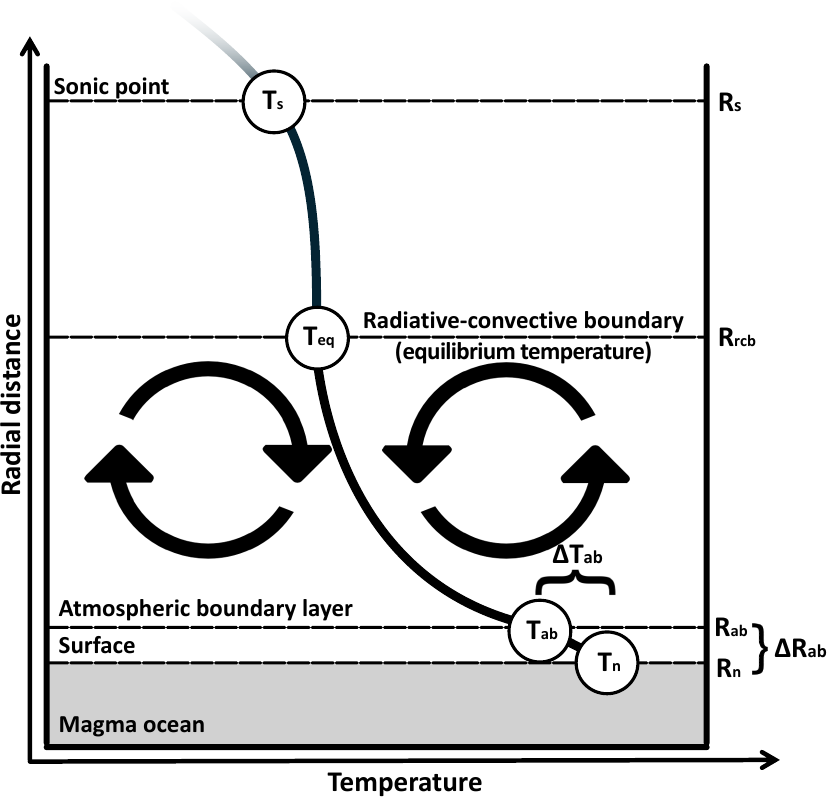}
    \caption{Schematic diagram showing the atmospheric structure adopted in this paper, which is based on the standard configuration used in the core powered mass loss model \citep{Ginzburg2016,Ginzburg2018,Gupta2018,Biersteker2019,Gupta2020}. Temperature and distance are not to scale; $T_{\rm s}$, $T_{\rm eq}$, $T_{\rm ab}$, and $T_{\rm n}$ are the temperature of the sonic point, the radiative-convective boundary (approximated as the equilibrium temperature), the top of the atmospheric boundary layer, and the nucleus surface. The boundary layer thickness and temperature contrast are denoted by $\Delta R_{\rm ab}$ and $\Delta T_{\rm ab}$, respectively. The core powered mass loss framework does not incorporate a boundary layer analysis but it is included here because it constitutes an important part of our modeling. In addition, whereas the standard model assumes an isothermal upper atmosphere \citep[e.g.,][]{Ginzburg2018}, our analysis accounts for cooling from advection. The temperature profile of the magma ocean is not shown because it is not directly required to model the atmospheric outflow; however, the thermal state of the magma ocean is ultimately what powers core powered mass loss and sets the timescale over which it operates.}
    \label{fig:schematic}
\end{figure}

The planetary model (Figure~\ref{fig:schematic}) consists of a hot interior with a magma ocean at temperature $T_{\rm n}$, and a hydrogen-rich atmosphere divided into two layers: a lower convective layer and an upper radiative layer. Whereas heating from the magma ocean drives atmospheric convection, stellar heating creates buoyant upper regions in which convection is suppressed. Because stellar heating affects one side of the planet more strongly than the other, it is challenging to precisely determine the boundary between radiative and convective zones. Without spherical symmetry, there is no simple prescription to establish where convection ceases and radiative heat transfer becomes dominant \citep[e.g., see the atmospheric modeling employed in;][]{Turbet2021,Selsis2023}. These complexities are avoided by treating the radiative section isothermal with temperature equal to the equilibrium temperature $T_{\rm eq}$. The radiative-convective boundary is therefore set by the location where the adiabatic temperature profile coincides with the equilibrium temperature. However, as we discuss later, in a hydrodynamic atmosphere the isothermal approximation is invalid at high altitudes where rapid gas advection leads to significant cooling.

Atmospheric escape becomes hydrodynamic when the sonic point (subscript s), where gas velocity equals the sound speed, is below the exobase, where particle collisions are rare and their trajectories are ballistic \citep[][Chapter 4]{walker1977}. Above the exobase, particles stop experiencing the collisional thrust force necessary to induce further outward acceleration. Without this force, they decelerate and gradually merge into the interplanetary medium. Thus, if the exobase were below the sonic point, gas cannot become transonic and mass loss is well approximated as being quasistatic, which is often modeled with Jeans escape \citep[][section 5.6]{Catling2017}. The transition between hydrodynamic outflow and Jeans escape is therefore governed by the collisional state of the gas, which in turn depends on the planet’s thermal and gravitational conditions. At high temperatures or for less gravitationally bound envelopes, collisions are frequent and a coherent, fluid-like outflow develops. In contrast, for cooler or more tightly bound atmospheres, escape occurs through the random motions of individual particles and the hydrodynamic wind model becomes invalid. A physically consistent and steady-state atmospheric outflow generally requires the gas velocity to pass through a sonic point to ensure that atmospheric density asymptotically approaches zero at large distances. Without a sonic transition, the atmosphere would remain subsonic and retain finite density at infinite radius, resulting in an unphysical atmosphere of infinite mass \citep{Parker1958,Parker1964}. This constraint has been confirmed by numerous numerical studies employing kinetic approaches such as the Direct Simulation Monte Carlo method \citep{Volkov2011,Volkov2013}, which consistently show that even flows initially subsonic near the exobase ultimately become supersonic because gas rarefaction at higher altitudes naturally accelerates gas particles outward.

The mass conservation equation quantifies the mass flux at any radius, but it is most naturally evaluated at the sonic point, where the advection speed equals the local sound speed ($c_{\rm s}{=}\sqrt{\gamma k_{\rm B}T_{\rm s}/\mu_{\rm s}}$, where $\gamma$, $k_{\rm B}$, $T_{\rm s}$, and $\mu_{\rm s}$ are the heat capacity ratio, Boltzmann constant, sonic point temperature, and mean molecular weight). This gives,
\begin{equation}
    \left.\frac{{\rm d}M_{\rm atm}}{{\rm d}t}\right|_{\rm GI{,}2} = 4 \pi R_{\rm s}^{2} \rho_{\rm s} c_{\rm s}.
\label{eq:GI_EL2}
\end{equation}
where $R_{\rm s}$ and $\rho_{\rm s}$ are the radius and density at the sonic point. The relevant measure of collisionality at this location is the Knudsen number, ${\rm Kn}_{\rm s} {\equiv} l_{\rm s}/H_{\rm s}$, where $l_{\rm s}$ is the mean free path and $H_{\rm s}$ is the local scale height. Here, the mean free path is given by $l_{\rm s} {=} (\sqrt{2}\pi d^2 n_{\rm s})^{-1}$, and the scale height is $H_{\rm s} {=} k_{\rm B} T_{\rm s} R_{\rm s}^2/\left(G M_{\rm n} \mu_{\rm s}\right)$, where $d$ is the molecular kinetic diameter, and $n_{\rm s}$ the number density at the sonic point. Hydrodynamic escape requires ${\rm Kn}_{\rm s} {<} 1$. When ${\rm Kn}_{\rm s} {>} 1$, the flow becomes collisionless before reaching the sonic point, and escape transitions to the Jeans regime. Setting ${\rm Kn}_{\rm s} {=} 1$ and solving for $n_{\rm s}$ yields,
\begin{equation}
n_{\rm s} = \frac{G M_{\rm n} \mu_{\rm s}}{\sqrt{2} \pi d^2 k_{\rm B} T_{\rm s} R_{\rm s}^2}.
\end{equation}
Combining with mass conservation, the critical mass loss rate below which escape is Jeans limited is determined,
\begin{equation}
    \left.\frac{{\rm d}M_{\rm atm}}{{\rm d}t}\right|_{\rm crit} = \frac{2\gamma\sqrt{2}}{d^{2}} \frac{GM_{\rm p}\mu_{\rm s}}{c_{\rm s}}.
\end{equation}
For characteristic planetary values, this is bounded from above by $\left.{\rm d}M_{\rm atm}/{\rm d}t \right|_{\rm crit} {\approx} 5{\times}10^{5}~{\rm kg~s^{-1}}$. Therefore, to ensure the validity of the core powered mass loss framework, we exclude from our model any cases where the predicted mass loss rate is below this threshold. The results presented here are thus exclusively in the hydrodynamic regime, where the model assumptions are valid.

The core powered mass loss framework applies to a planet immediately following a giant impact after the protoplanetary disk has dissipated. If the protoplanetary disk were still present, a planet would reaccrete a new hydrogen-rich atmosphere after a giant impact. The planetary nucleus is hot and largely molten, with the atmosphere above having a temperature profile that is inherently unstable and hydrodynamic. Because of mass conservation, the mass flux can be determined at any radial distance (see Equation~\ref{eq:GI_EL2}). The point of contention lies with the other mass loss equation in the standard core powered mass loss framework: the ``energy limited'' constraint (Equation~\ref{eq:GI_EL1}). The argument is that the mass loss rate is limited by the luminosity through the radiative-convective boundary. The maximum mass loss rate occurs when all internal luminosity is converted into work against gravity, setting an energy limit. Therefore, the mass loss rate is suggested to be the minimum of the ``energy limited'' (Equations~\ref{eq:GI_EL1}) and mass conservation equations (Equation~\ref{eq:GI_EL2}),
\begin{equation}
    \left.\frac{{\rm d}M_{\rm atm}}{{\rm d}t}\right|_{\rm GI} = \min{\left(\left.\frac{{\rm d}M_{\rm atm}}{{\rm d}t}\right|_{\rm GI{,}1}, \left.\frac{{\rm d}M_{\rm atm}}{{\rm d}t}\right|_{\rm GI{,}2}\right)}.
\label{eq:mass_loss_final}
\end{equation}

This argument would be reasonable if mass loss should not exceed the luminosity escaping from the planetary interior. In the next section, we examine the assumptions behind the ``energy limited'' argument and show that mass loss is actually more complicated than what Equation~\ref{eq:GI_EL1} suggests.

\section{On the ``energy limited'' argument}
\label{sec:assumptions}

The ``energy limited'' argument states that mass loss is controlled by the luminosity through the radiative-convective boundary, with planetary cooling occurring proportionally through this luminosity. Before presenting its formal analysis, it is worth considering a simple analogy: a sealed gas cylinder placed on a table, where the gas in the cylinder represents the planetary atmosphere and the table represents the planetary surface and interior. When the gas is cold and mildly pressurized, opening the valve produces a low mass flux; the gas expands and cools adiabatically, but the driving pressure differential is small. When the gas has first been heated so that it is highly pressurized, opening the valve produces a high mass flux; the gas again expands and cools adiabatically, but the large pressure differential drives a much stronger outflow. In both cold and hot cases above, the outflow is powered by energy originally stored in the gas itself, and it is incorrect to assume that the table beneath the cylinder also has to lose energy at the same rate. The core powered mass loss framework makes an analogous error by assuming that, as atmospheric mass is lost, the planetary interior must continuously cool to provide the energy required for escape. Our analysis demonstrates that, just as in the gas cylinder analogy, after a giant impact heats the planet’s interior, the atmosphere assumes an inherently unstable high-temperature profile. This instability alone drives a hydrodynamic outflow without requiring ongoing energy from the planet’s interior. In the following sections we provide quantitative arguments to show that a planet experiencing mass loss is indeed analogous to the heated gas cylinder. In section~\ref{sec:interior_cooling}, we show that the planetary interior does not cool and remains mostly unaffected during atmospheric loss. In section~\ref{sec:atmospheric_cooling}, we show that the atmosphere expands adiabatically when mass is lost and that it can therefore not be modeled as being isothermal.

\subsection{The planetary interior does not cool from atmospheric loss}
\label{sec:interior_cooling}

In the analogy of the gas cylinder, the table does not cool because the gas cylinder and the table do not have a mechanism through which they can exchange energy efficiently. In this section, we show that the same occurs for the atmosphere and the magma ocean. 

Gases expand adiabatically into space during atmospheric loss, resulting in a net outward energy flux. It may therefore be argued that catastrophic mass loss efficiently cools the entire planet. According to \citet{Owen2016}, the cooling rate from mass loss is,
\begin{equation}
    L_{\rm m.l.} = \frac{\gamma}{\gamma - 1} \dot{M}_{\rm atm} c_{\rm s}^{2},
\label{eq:luminosity_mass_loss}
\end{equation}
where $\gamma$ is the heat capacity ratio, $\dot{M}_{\rm atm}$ is the mass loss rate, and $c_{\rm s}$ is the isothermal sound speed. In this paper, we adopt $\gamma{=}5/3$, corresponding to an endmember case of a fully neutral or fully ionized hydrogen–helium gas. This value is used to generate our figures, but the underlying theory applies for all $\gamma$ relevant to hydrodynamic planetary atmospheres undergoing mass loss through the core powered mechanism. For atomic hydrogen,
\begin{equation}
    L_{\text{m.l.}} = 2.06 \times 10^{22} \left(\frac{\dot{M}_{\rm atm}}{10^{15}\, \mathrm{{kg/s}}}\right) \left( \frac{{T_s}}{{10^{3}\, \mathrm{{K}}}} \right) \, \mathrm{{J/s}}
\label{eq:luminosity_mass_loss_2}
\end{equation}
with $T_{\rm s}$ being the temperature at the sonic point. However, this reasoning is valid only when there is efficient energy transfer through the planet. If energy transfer is inefficient between the planetary interior and atmosphere, the atmosphere will decompress and assume its new equilibrium thermal structure while the interior remains virtually unaffected \citep{Modirrousta2023}. Determining whether such cooling occurs is necessary to evaluate whether a planet can remain sufficiently hot to maintain an extreme mass loss. To address this question, we consider a planet with very high internal temperatures so that its temperature profile is well approximated as being isothermal because the sonic point is close to the planetary surface. Atmospheric pressure decreases with mass loss, reducing the internal atmospheric temperature and creating a larger temperature contrast across the boundary layer between the planetary surface and atmosphere. This amplifies energy transfer through the boundary layer, which is governed by conduction,
\begin{equation}
    F_{\rm ab} = \frac{k \rho c_{P}\Delta T_{\rm ab}}{\Delta R_{\rm ab}},
\label{eq:conductive_flux}
\end{equation}
with $k$ being the thermal diffusivity, $\rho$ the density, $c_{P}$ the specific heat capacity, $\Delta T_{\rm ab}$ the temperature contrast, and $\Delta R_{\rm ab}$ the boundary layer thickness, which is determined by the critical Rayleigh number,
\begin{equation}
    {\rm Ra_{c}} = \frac{\rho g \alpha \Delta T_{\rm ab} \Delta R_{\rm ab}^{3}}{\eta k}.
\label{eq:Rayleigh_number}
\end{equation}
Here, $g$ is the gravitational acceleration, $\alpha$ is the thermal expansion coefficient, and $\eta$ is the viscosity. Combining Equations~\ref{eq:conductive_flux} and \ref{eq:Rayleigh_number},
\begin{equation}
    F_{\rm ab} = k^{\frac{2}{3}} \rho^{\frac{4}{3}} c_{P} \Delta T_{\rm ab}^{\frac{4}{3}} \left(\frac{g \alpha }{\eta {\rm Ra_{c}}}\right)^{\frac{1}{3}}.
\label{eq:F_b}
\end{equation}
From Chapman-Enskog theory \citep{Chapman1970},
\begin{equation}
    k = \frac{75}{64d^{2}} \frac{k_{\rm B}}{\rho c_{P}\bar{\mu}} \left(\frac{\bar{\mu} k_{\rm B}T}{\pi}\right)^{\frac{1}{2}},
\label{eq:kappa}
\end{equation}
and
\begin{equation}
    \eta = \frac{5}{16d^{2}} \left(\frac{\bar{\mu} k_{\rm B}T}{\pi}\right)^{\frac{1}{2}},
\label{eq:eta}
\end{equation}
where $d$ is the kinetic diameter. Combining with Equation~\ref{eq:F_b},
\begin{equation}
    F_{\rm ab} = \left(\frac{1125}{256 \pi^{\frac{1}{2}}}\right)^{\frac{1}{3}} d^{-\frac{2}{3}} k_{\rm B}^{\frac{5}{6}} \bar{\mu}^{-\frac{1}{2}} \rho^{\frac{2}{3}} c_{\rm p}^{\frac{1}{3}} \Delta T_{\rm ab}^{\frac{4}{3}} g^{\frac{1}{3}} \alpha^{\frac{1}{3}} {\rm Ra_{c}}^{-\frac{1}{3}} T^{\frac{1}{6}}.
\end{equation}
Recognizing that $c_{\rm p} {=} \gamma/\left(\gamma-1\right)k_{\rm B}/\bar{\mu}$, $\alpha{=}1/T$, $g{=}g_{\oplus}\left(M_{\rm n}/M_{\oplus}\right)^{1/2}$ \citep[i.e., $R_{\rm n}{\propto}M_{\rm n}^{1/4}$;][]{Zeng2016}, and multiplying by the surface area to get the luminosity,
\begin{equation}
    L_{\rm ab} = 4 \pi R_{\oplus}^{2} \left(\frac{1125 g_{\oplus}}{256 \pi^{\frac{1}{2}}}\right)^{\frac{1}{3}} \left(\frac{\gamma}{\gamma-1}\right)^{\frac{1}{3}} d^{-\frac{2}{3}} k_{\rm B}^{\frac{7}{6}} \bar{\mu}^{-\frac{5}{6}} \rho^{\frac{2}{3}} \Delta T_{\rm ab}^{\frac{4}{3}} \left(\frac{M_{\rm n}}{M_{\oplus}}\right)^{\frac{2}{3}} T^{-\frac{1}{6}} {\rm Ra_{c}}^{-\frac{1}{3}}.
\end{equation}
Using $d{=}d_{\rm H}{=}2.2{\times}10^{-10}~{\rm m}$, $\bar{\mu}{=}\mu_{\rm H}{=}1.674{\times}10^{-27}~{\rm kg}$, $\gamma{=}\gamma_{\rm H}{=}5/3$, and ${\rm Ra_{c}}{\approx}1000$,
\begin{equation}
    L_{\rm ab} = 1.12 \times 10^{21} \left(\frac{\rho}{10^{2}\, \mathrm{{kg/m^{3}}}}\right)^{\frac{2}{3}} \left(\frac{\Delta T_{\rm ab}}{{10^{3}\, \mathrm{{K}}}}\right)^{\frac{4}{3}} \left(\frac{T}{{10^{4}\, \mathrm{{K}}}}\right)^{-\frac{1}{6}} \left(\frac{M_{\rm n}}{M_{\oplus}}\right)^{\frac{2}{3}} \, \mathrm{{J/s}}.
\end{equation}
We consider the limiting case when $\Delta T_{\rm ab} {=} T$, corresponding to the maximum possible flux through the boundary layer, 
\begin{equation}
    \mathcal{O}(L_{\rm ab}) = 2.42 \times 10^{22} \left(\frac{\rho}{10^{2}\, \mathrm{{kg/m^{3}}}}\right)^{\frac{2}{3}} \left(\frac{T}{{10^{4}\, \mathrm{{K}}}}\right)^{\frac{7}{6}} \left(\frac{M_{\rm n}}{M_{\oplus}}\right)^{\frac{2}{3}}  \, \mathrm{{J/s}}.
\label{eq:luminosity_boundary}
\end{equation}
Comparing Equations~\ref{eq:luminosity_mass_loss_2} and \ref{eq:luminosity_boundary}, in the maximally hydrodynamic limit where the sonic point lies close to the planetary surface ($T_{\rm s}{\to}T$),
\begin{equation}
    \frac{\mathcal{O}(L_{\rm ab})}{L_{\rm m.l.}} = 0.12 \left(\frac{\rho}{10^{2}\, \mathrm{{kg/m^{3}}}}\right)^{\frac{2}{3}}\left(\frac{\dot{M}_{\rm atm}}{10^{15}\, \mathrm{{kg/s}}}\right)^{-1} \left(\frac{T}{{10^{4}\, \mathrm{{K}}}}\right)^{\frac{1}{6}} \left(\frac{M_{\rm n}}{M_{\oplus}}\right)^{\frac{2}{3}}.
\end{equation}
Using characteristic values of $\rho{\sim}10^{2}{\rm ~kg~m^{-3}}$, $\dot{M}_{\rm atm}{\sim}10^{15}{\rm ~kg~s^{-1}}$, $T{\sim}10^{4}~{\rm K}$, and $M{=}3~M_{\oplus}$, one gets $\mathcal{O}(L_{\rm ab})/L_{\rm m.l.}{\sim}10^{-1}$. This demonstrates that even in the limiting case when $\Delta T_{\rm ab} {=} T$, energy transfer across the boundary layer is insufficient to compensate for the energy lost through mass loss. Given that we are considering the maximum temperature contrast and that we focus on planets with small atmospheres relative to their total masses, it is safe to conclude that the planetary interior remains unaffected by mass loss. Like in the gas cylinder analogy, there is no efficient energy transfer mechanism between the atmosphere and the magma ocean.

\subsection{Atmospheric gas must cool below the sonic point}
\label{sec:atmospheric_cooling}

In this section, we show that the potential energy of the atmosphere drives gas advection. Escaping gas will cool adiabatically when it expands and rises. The standard core powered mass loss framework proposes that this cooling is offset by energy supplied from the cooling magma ocean to sustain an isothermal outer atmosphere. However, as gas ascends, it does work against gravity by lifting its own weight. This work is powered by the internal energy of the gas, which leads to a reduction in temperature. Maintaining an isothermal profile under these conditions therefore requires a continuous supply of energy. Whereas the total internal luminosity is indeed constant with radius, the radiative flux decreases with altitude because of the inverse square law. At the same time, radial velocity increases with altitude, increasing the magnitude of adiabatic cooling. This creates a mismatch between the rate of energy loss and the energy available to compensate for it. In the absence of a mechanism capable of delivering sufficient energy to the outer layers, the atmosphere cannot remain isothermal and must assume a negative temperature gradient with radial distance. Indeed, recent radiation-hydrodynamic modeling by \citet{Misener2025} also indicates deviations from isothermality above the radiative-convective boundary. It follows that the temperature gradient above the radiative-convective boundary in a hydrodynamic atmosphere is,
\begin{equation}
    \frac{{\rm d}T}{{\rm d}r} = - \Phi_{\rm int}\left(r\right) - \Phi_{\rm ext}\left(r\right) - \Phi_{\rm mix}\left(r\right) - \frac{\gamma -1}{\gamma} \frac{\bar{\mu} v}{k_{\rm B}}\frac{{\rm d}v}{{\rm d}r},
\label{eq:temperature_gradient_1}
\end{equation}
where $\Phi_{\rm int}(r)$ represents internal heating, $\Phi_{\rm ext}(r)$ accounts for external heating, $\Phi_{\rm mix}(r)$ describes mixing processes such as eddy diffusion, which drive the system toward an adiabatic temperature profile, and the final term represents adiabatic cooling caused by radial advection (see Appendix~\ref{sec:dT_dr} for its derivation). Here, $T$ is temperature, $\gamma$ is the heat capacity ratio, and $r$ is the radial distance. Following the core powered mass loss framework, we assume that internal heating, external heating, and mixing approximately cancel out, resulting in a nearly isothermal upper atmosphere. Deviations from isothermality arise primarily from the last term, i.e., adiabatic cooling by gas advection. Near the radiative-convective boundary, where velocities and velocity gradients are small, this term is negligible, and the temperature remains nearly constant. In the upper atmosphere, however, where velocity and acceleration increase, the advection term becomes significant. The temperature gradient then simplifies to,
\begin{equation}
    \frac{{\rm d}T}{{\rm d}r} = - \frac{\gamma - 1}{\gamma} \frac{\bar{\mu} v}{k_{\rm B}} \frac{{\rm d}v}{{\rm d}r}.
\label{eq:temperature_gradient_3}
\end{equation}
Integrating from the radiative-convective boundary to the sonic point yields (see Appendix~\ref{sec:dT_dr} for its derivation),
\begin{equation}
\frac{T_{\rm s}}{T_{\rm rcb}} = \frac{2}{\gamma+1 - \left(\gamma-1\right)\left(\frac{v_{\rm rcb}}{c_{\rm s}}\right)^{2}}.
\label{eq:T_ratio}
\end{equation}
Convection in the region below the radiative-convective boundary acts to homogenize entropy and rapidly erase any net radial flow. Mass loss therefore proceeds as a uniform contraction of the convective region, rather than through localized outflow. In steady state, the mean radial velocity at this boundary has to be close to zero because it separates the well-mixed, hydrostatic convective layer from the hydrodynamic wind above. Therefore, $v_{\rm rcb} {\approx} 0$, yielding, $T_{\rm s}/T_{\rm rcb} {=} 2/\left(\gamma {+} 1\right)$, and for $\gamma {=} 5/3$, this evaluates to $T_{\rm s}/T_{\rm rcb} {=} 0.75$. Thus, under realistic atmospheric conditions, the upper atmosphere cannot be modeled as strictly isothermal.

The above derivation assumes that a steady-state outflow exists. However, whether such a solution can be sustained remains uncertain. Whereas a purely adiabatic outflow can maintain a steady-state temperature profile without any radiative transport, the structure derived here is not purely adiabatic. Near the radiative-convective boundary, the temperature remains nearly isothermal because of radiative diffusion, and the profile transitions gradually as advection becomes more important with altitude. For this temperature structure to be maintained in steady state, the energy lost to adiabatic expansion must be resupplied fast enough to keep the temperature fixed at each radius. This resupply may occur through a combination of radiative diffusion and advective energy transport from below. Our timescale comparison evaluates whether radiative diffusion is sufficiently rapid to avoid being a limiting factor, that is, whether the system would enter an ``energy limited'' regime.

\begin{figure}
    \centering
    \includegraphics[width=0.8\linewidth]{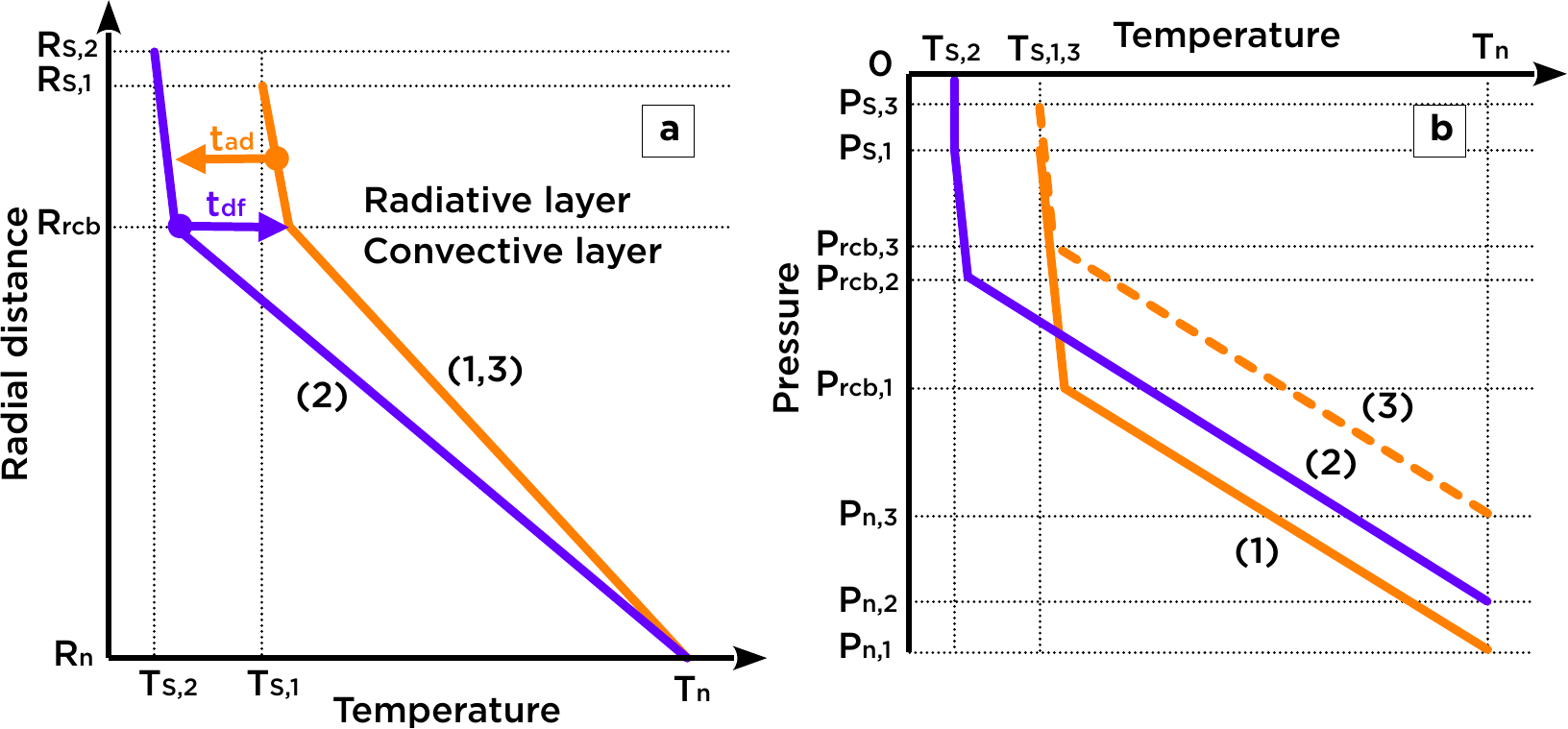}
    \caption{Schematic diagram showing the atmospheric temperature profile with respect to (a) radial distance and (b) pressure, before (orange; subscript 1) and immediately after (blue; subscript 2) experiencing mass loss. The orange dashed line (subscript 3) represents the profile after radiative heating dominates over advective cooling. The advection and diffusion timescales are $t_{\rm ad}$ and $t_{\rm df}$, respectively. Radii $R_{\rm s}$, $R_{\rm rcb}$, and $R_{\rm n}$ correspond to the sonic point, the radiative-convective boundary, and the surface of the planetary nucleus, respectively. The temperature of the sonic point is $T_{\rm s}$ and the temperature of the planetary nucleus is $T_{\rm n}$. The sonic point of profile 2 is located at a higher elevation because it is inversely proportional to temperature. The radius of the radiative-convective boundary is mostly independent of mass loss because, for an irradiated planet, it is thought to scale with the ratio of the incoming radiant flux to the internal heat flux. This ratio is unlikely to change when mass is lost. Profile 2 is unlikely to be realized because radiative heating dominates over advective cooling within the atmosphere (i.e., $t_{\rm df}{<}t_{\rm ad}$). The atmosphere will transition directly from profile 1 to 3 instead. Pressure profile 3 is smaller than 2 because the atmosphere adiabatically expands to restore the lost mass at the sonic point.}
    \label{fig:cooling_cartoon}
\end{figure}

Figure~\ref{fig:cooling_cartoon} is a schematic diagram showing the temperature profile of an atmosphere before (1) and after (2) experiencing mass loss. 
The advection timescale is given by,
\begin{equation}
    t_{\rm ad} = \frac{l}{c_{\rm s}},
\end{equation}
and the radiative diffusion timescale by,
\begin{equation}
    t_{\rm df} = \frac{l^{2}}{2k_{\rm d}},
\end{equation}
where $l$ is an arbitrary length scale, $c_{\rm s}$ is the sound speed, and $k_{\rm d}$ is the radiative diffusivity. When $t_{\rm df}{<}t_{\rm ad}$, radiative diffusion operates quickly enough to replace the energy lost by adiabatic expansion, ensuring that it does not become an energy bottleneck in maintaining a steady-state temperature at each radius. The radiative diffusivity is $k_{\rm d}{=}v_{\lambda}{/}\left(3\rho_{\rm s}\bar{\kappa}\right)$, where $v_{\lambda}$ is the light speed, $\rho_{\rm s}$ is the sonic point density, and $\bar{\kappa}$ is the mean opacity. Combining together we obtain,
\begin{equation}
    l < \frac{2}{3} \frac{v_{\lambda}}{\rho_{\rm s} \bar{\kappa} c_{\rm s}}.
\end{equation}
Multiplying across by $\rho_{\rm s}\bar{\kappa}$ and approximating the optical depth across the length scale as $\tau{\sim}\rho_{\rm s}\bar{\kappa}l$ yields,
\begin{equation}
    \tau \lesssim \frac{2}{3} \frac{v_{\lambda}}{c_{\rm s}}.
\end{equation}
For $c_{\rm s} {\sim} 2000~{\rm m~s^{-1}}$, radiative diffusion operates quickly enough in regions with $\tau {<} 10^5$ to avoid becoming a bottleneck for maintaining a steady-state temperature profile. We can compare this value to the optical depth at the sonic point, which is set by the blanketing effect of the outgoing radial winds \citep[see Equation~22 of][]{Modirrousta2023}:
\begin{equation}
    \tau_{\rm s} = \frac{2}{\gamma} \bar{\kappa} \rho_{\rm s}R_{\rm s}
\end{equation}
Using characteristic values of $\bar{\kappa}{\sim}1~{\rm m^{2}~kg^{-1}}$, $\rho_{\rm s}{\sim}10^{-5}~{\rm kg~m^{-3}}$, and $R_{\rm s}{\sim}10^{8}~{\rm m}$ we obtain $\tau_{\rm s}{\sim}10^{3}$. Using smaller opacities \citep[e.g., $\bar{\kappa}{\sim}10^{-2}~{\rm m^{2}~kg^{-1}}$;][]{Freedman2014} yields even lower optical depths at the sonic point. This timescale comparison demonstrates that radiative diffusion acts quickly enough in regions below the sonic point to compensate for energy lost by adiabatic expansion, ensuring that the temperature at each radius remains fixed in time and a steady state solution exists. Moreover, this finding supports our earlier conclusion that the mass loss rate is not constrained by internal luminosity through the radiative-convective boundary.

We have thus shown that the gravitational potential energy of the atmosphere drives gas advection, and that below the sonic point, the adiabatic cooling can be sustained in steady state through a combination of radiative diffusion and upward energy transport from deeper layers. Our analysis demonstrates that radiative diffusion operates quickly enough not to impede this balance, ensuring that the temperature at each radius can remain fixed in time.

\section{The revised model}
\label{sec:outlook}

In the previous section, we showed that the gas advection is not powered by the luminosity through the radiative-convective boundary, but by the potential energy of the atmosphere, which expands adiabatically to do work against gravity. This renewed understanding allows us to refine the description of the core powered mass loss mechanism: A planet hosting a primordial hydrogen-rich atmosphere experiences a giant impact that significantly increases its mantle temperature and causes its atmosphere to establish a new temperature profile. At sufficiently high temperatures, the atmosphere is intrinsically unstable, and gas is too hot for the planet to retain the atmosphere, leading to hydrodynamic escape at the sonic point. At the sonic point, local hydrostatic equilibrium is disrupted because the internal gas pressure exceeds the weight of the overlying atmosphere. This imbalance induces an upward acceleration, causing gas from below to advect adiabatically. The process does not cool the magma ocean but it does cool gases near the sonic point where gas advection is significant. Equation~\ref{eq:mass_loss_final} should therefore be changed to,
\begin{equation}
    \left.\frac{{\rm d}M_{\rm atm}}{{\rm d}t}\right|_{\rm GI} = \left.\frac{{\rm d}M_{\rm atm}}{{\rm d}t}\right|_{\rm GI{,}2} = 4 \pi R_{\rm s}^{2} \rho_{\rm s} c_{\rm s},
\label{eq:mass_loss_final2}
\end{equation}
with mass loss being set simply by mass conservation. Whereas this approach is more involved than the standard ``energy limited'' model, it can still yield tractable approximations, as shown in Section~\ref{sec:approximation}.

We now provide a quantitative comparison between Equations~\ref{eq:mass_loss_final} and \ref{eq:mass_loss_final2}. To model gas outflow, we use the Euler equations with mass and momentum conservation given by,
\begin{equation}
    \frac{{\rm d}\left[\ln{\left(n v r^{2}\right)}\right]}{{\rm d}r} = \frac{1}{n}\frac{{\rm d}n}{{\rm d}r} + \frac{1}{v}\frac{{\rm d}v}{{\rm d}r} + \frac{2}{r} = 0,
\label{eq:mass_conservation}
\end{equation}
and
\begin{equation}
    \frac{{\rm d}P}{{\rm d}r} = -n\bar{\mu} g - n\bar{\mu} v \frac{{\rm d}v}{{\rm d}r},
\label{eq:momentum_conservation}
\end{equation}
respectively, where $n$ is the volume particle number density, $v$ is velocity, $r$ is radial distance, $P$ is pressure, $\bar{\mu}$ is the mean molecular weight of gas, and $g$ is the gravitational acceleration. From the ideal gas equation, and letting $k_{\rm B}$ and $T$ be the Boltzmann constant and temperature,
\begin{equation}
    \frac{{\rm d}P}{{\rm d}r} = k_{\rm B}T \frac{{\rm d}n}{{\rm d}r}+nk_{\rm B} \frac{{\rm d}T}{{\rm d}r},
\end{equation}
which upon combining with Equation~\ref{eq:momentum_conservation} yields,
\begin{equation}
    \frac{{\rm d}n}{{\rm d}r} = -\frac{n\bar{\mu} g}{k_{\rm B}T} - \frac{n\bar{\mu} v}{k_{\rm B}T} \frac{{\rm d}v}{{\rm d}r} - \frac{n}{T}\frac{{\rm d}T}{{\rm d}r}.
\label{eq:dn_dr}
\end{equation}
Equation~\ref{eq:dn_dr} is then combined with mass conservation to give,
\begin{equation}
    \left(v^{2}-\frac{k_{\rm B}T}{\bar{\mu}}\right)\frac{{\rm d}v}{{\rm d}r} = \frac{2k_{\rm B}Tv}{\bar{\mu} r^{2}}\left(r-\frac{GM_{\rm n}\bar{\mu}}{2k_{\rm B}T}-\frac{r^{2}}{2T}\frac{{\rm d}T}{{\rm d}r}\right).
\label{eq:revealing_equation}
\end{equation}
Keeping in line with the standard core powered mass loss model, which treats the upper atmosphere isothermal, we assume deviations from isothermality occur only from gas advection. By combining Equations~\ref{eq:temperature_gradient_3} and \ref{eq:revealing_equation} and then rearranging the terms, we obtain
\begin{equation}
    \left(\frac{v\bar{\mu}}{\gamma k_{\rm B}T}-\frac{1}{v}\right)\frac{{\rm d}v}{{\rm d}r} = \frac{2}{r} - \frac{GM_{\rm n}\bar{\mu}}{k_{\rm B}Tr^{2}},
\label{eq:revealing_equation_2}
\end{equation}
from which the sonic point naturally emerges as the radius where $r {=} R_{\rm s} {=} GM_{\rm n}\bar{\mu}/(2k_{\rm B}T)$ and $v {=} v_{\rm s} {=} \sqrt{\gamma k_{\rm B}T/\bar{\mu}}$. To eliminate the temperature dependence on the left-hand side, we integrate Equation~\ref{eq:temperature_gradient_3} between the limits $T$ and $T_{\rm s}$ for temperature and $v$ and $v_{\rm s}$ for velocity. This integration yields $\gamma k_{\rm B}T/\bar{\mu}{=}\left(\gamma{+}1\right)v_{\rm s}^{2}/{2}{-}\left(\gamma{-}1\right)v^{2}/2$, which is then substituted back into Equation~\ref{eq:revealing_equation_2} to obtain,
\begin{equation}
    \left[\frac{2v}{\left(\gamma{+}1\right)v_{\rm s}^{2}{-}\left(\gamma{-}1\right)v^{2}}-\frac{1}{v}\right]\frac{{\rm d}v}{{\rm d}r} = \frac{2}{r} - \frac{GM_{\rm n}\bar{\mu}}{k_{\rm B}Tr^{2}}.
\label{eq:revealing_equation_3}
\end{equation}
The left-hand side is integrable, but we cannot integrate the right-hand side because the relation between $T$ and $r$ is unknown. Several numerical approaches have been developed to solve such wind problems. \citet{Murray2009} use a relaxation method that iteratively converges on a transonic solution by enforcing boundary conditions at the base of their simulation grid and the sonic point. This method is sensitive to the choice of boundary conditions; for example, a $20\%$ percent variation in the base radius can lead to a systematic uncertainty in the mass-loss rate of up to a factor of two. \citet{Garcia2007} employs a different strategy based on shooting methods, in which one guesses initial conditions and integrates outward, adjusting the guess until the outer boundary condition is satisfied. Because solutions can diverge near the sonic point, artificial dissipation is added to smooth the transition and ensure numerical stability. This dissipation can affect the accuracy of the velocity gradient, and its impact is difficult to quantify without detailed convergence tests. Whereas these methods were applied in the context of X-ray and ultraviolet induced photoevaporation, the underlying numerical challenges are similar to those relevant to core powered mass loss.

In our case, the wind temperature is monotonic and well constrained by boundary conditions. As shown in Section~\ref{sec:atmospheric_cooling}, the temperature gradient is small below the sonic point, with the temperature at the sonic point differing from its value at the radiative convective boundary by ${\sim}25\%$. This indicates that the gas is close to isothermal throughout the region of interest. Furthermore, because the mass-loss rate is determined by conditions at the sonic point, variations in temperature at larger radii do not affect our results. For these reasons, we prescribe a parsimonious, physically motivated function for $T(r)$ and integrate Equation~\ref{eq:revealing_equation_3} directly, avoiding the numerical complications associated with the sonic point while maintaining the accuracy required for modeling core powered mass loss. We therefore seek a functional form for $T(r)$ that satisfies the boundary conditions. First, Equation~\ref{eq:T_ratio} suggests $T(r{\to}R_{\rm rcb})/T_{\rm s}{\approx} (\gamma+1)/2$ (for $R_{\rm rcb}{\ll}R_{\rm s}$). Next, Equation~\ref{eq:revealing_equation_3} requires that the velocity reaches a maximum of $v(r{\to}\infty)/v_{\rm s} {=} \sqrt{(\gamma+1)/(\gamma-1)}$, which, when combined with Equation~\ref{eq:temperature_gradient_3}, yields $T(r{\to}\infty)/T_{\rm s} {\to} 0$. Last, to ensure the wind reaches a finite terminal velocity as gravitational and pressure forces vanish, we require ${\rm d}v/{\rm d}r {\to} 0$ as $r {\to} \infty$. For this condition to be met, the right-hand side of Equation~\ref{eq:revealing_equation_3} must vanish, which occurs only if $T(r) {\propto} 1/r$ at large $r$. As we show in Appendix~\ref{sec:unique}, this asymptotic behavior implies that ${\rm d}T/{\rm d}r {\propto} {-}T^2$ to leading order. By adopting this relation as a global model and applying the appropriate boundary conditions, we obtain the unique analytic solution,
\begin{equation}
    \frac{T(r)}{T_{\rm s}} = \frac{\gamma+1}{2+\left(\gamma-1\right)\frac{r}{R_{\rm s}}},
\label{eq:T_solution}
\end{equation}
whose uniqueness is guaranteed by the Picard–Lindelöf theorem for the assumed global form of the temperature gradient, as demonstrated in Appendix~\ref{sec:unique}. There, we also show that the velocity profile between $R_{\rm rcb}{\leq}r{\leq}R_{\rm s}$ is insensitive to the choice of temperature profile by comparing Equation~\ref{eq:T_solution} with alternative trial functions. Substituting Equation~\ref{eq:T_solution} into Equation~\ref{eq:revealing_equation_3} and integrating from $R_{\rm s}$ to $r$ and from $v_{\rm s}$ to $v$ yields,
\begin{equation}
    \frac{1}{\gamma-1}\ln\left[\frac{\gamma+1}{2}-\frac{\gamma-1}{2}\left(\frac{v}{v_{\rm s}}\right)^{2}\right]+\ln\left(\frac{v}{v_{\rm s}}\right)=\frac{4}{\gamma+1}\ln\left(\frac{R_{\rm s}}{r}\right)+\frac{4}{\gamma+1}\left(1-\frac{R_{\rm s}}{r}\right).
\label{eq:v_sol}
\end{equation}

The temperature and density profiles for an ideal gas hydrostatic adiabatic interior atmosphere \citep[i.e., convecting; see][for full details]{Modirrousta2023} with a hydrodynamic outer atmosphere are,
\begin{equation}
    T(r) = \begin{cases}
  T_{\rm eq}\left[1 + \frac{\gamma-1}{\gamma}\frac{GM_{\rm n} \bar{\mu}}{k_{\rm B}T_{\rm eq}}\left(\frac{1}{r}-\frac{1}{R_{\rm rcb}}\right)\right]  & r \leq R_{\rm rcb} \\
  T_{\rm eq}\frac{2+\left(\gamma-1\right)\frac{R_{\rm rcb}}{R_{\rm s}}}{2+\left(\gamma-1\right)\frac{r}{R_{\rm s}}} & r > R_{\rm rcb}
\end{cases},
\label{eq:T_profile}
\end{equation}
and,
\begin{equation}
    \rho(r) = \begin{cases}
  \rho_{\rm rcb}\left[1 + \frac{\gamma-1}{\gamma}\frac{GM_{\rm n} \bar{\mu}}{k_{\rm B}T_{\rm eq}}\left(\frac{1}{r}-\frac{1}{R_{\rm rcb}}\right)\right]^{\frac{1}{\gamma-1}}  & r \leq R_{\rm rcb} \\
  \rho_{\rm rcb}\left(\frac{R_{\rm rcb}}{r}\right)^{2}\left(\frac{v_{\rm rcb}}{v}\right)  & r > R_{\rm rcb}
\end{cases},
\label{eq:rho_profile}
\end{equation}
respectively. The adiabatic temperature and density profiles are obtained by combining hydrostatic equilibrium, the ideal gas law, and the isentropic relations. The hydrodynamic temperature profile is a rescaling of Equation~\ref{eq:T_solution} whereas the density profile is set by mass conservation, with the velocity ratio $v_{\rm rcb}/v$ given by Equation~\ref{eq:v_sol}. The radiative-convective boundary radius $R_{\rm rcb}$ is estimated by solving the adiabatic section of Equation~\ref{eq:T_profile} with $r{=}R_{\rm n}$ and $T{=}T_{\rm n}$. For reference, the standard core powered mass loss model assumes an isothermal, hydrostatic upper atmosphere and uses the barometric formula to model its density structure \citep[Equation~5;][]{Biersteker2019}. The density profile is then combined with mass conservation to evaluate atmospheric loss. We will henceforth refer to this approach as the isothermal mass conservation equation, to distinguish it from our hydrodynamic treatment, which relaxes both the isothermal and hydrostatic assumptions.

The density of the radiative-convective boundary is,
\begin{equation}
    \rho_{\rm rcb} = \frac{3M_{\rm atm}}{4 \pi \int^{R_{\rm s}}_{\rm R_{\rm n}}r^{2}\rho(r)/\rho_{\rm rcb}{\rm d}r},
\label{eq:rho_rcb}
\end{equation}
where $\rho(r)$ is given by Equation~\ref{eq:rho_profile}, $R_{\rm n}$ is the radius of the planetary nucleus \citep[approximately $R_{\rm n}/R_{\oplus}{=}\left(M_{\rm n}/M_{\oplus}\right)^{1/4}$ for an Earth-like composition;][]{Zeng2013,Zeng2016}, and $M_{\rm atm}$ is the atmospheric mass of choice. The mass conservation equation (Equation~\ref{eq:GI_EL2}) can be evaluated using the above three equations.

For Equation~\ref{eq:GI_EL1}, the ``energy limited'' equation, the planetary luminosity is required. The standard core powered mass loss framework uses \citep{Ginzburg2018,Gupta2018,Biersteker2019,Gupta2020},
\begin{equation}
    L_{\rm int}=\frac{\gamma - 1}{\gamma} \frac{64 \pi \sigma T_{\rm rcb}^{3} G M_{\rm n} \bar{\mu}}{3 \bar{\kappa} \rho_{\rm rcb} k_{\rm B}},
\label{eq:luminosity}
\end{equation}
which is derived by combining the equation for radiative diffusion through a spherical system with the adiabatic temperature gradient. Here, $\sigma$ is the Stefan-Boltzmann constant and $T_{\rm rcb}{=}T_{\rm eq}$ (Figure~\ref{fig:schematic}). Whereas Equation~\ref{eq:luminosity} is valid for internally heated systems like the Sun, it is hard to justify for planets whose upper atmospheres are primarily heated by incoming stellar radiation because this radiation affects one side of the planet more strongly than the other, breaking spherical symmetry. Most importantly, Equation~\ref{eq:luminosity} does not isolate the incoming stellar luminosity component from the internal intrinsic component. A more accurate approach would use Equation~\ref{eq:temperature_gradient_1} with $\Phi_{\rm int}\left(r\right)$ and $\Phi_{\rm ext}\left(r\right)$ as functions of the internal and external luminosities respectively; the third term, $\Phi_{\rm mix}\left(r\right)$, would depend on the mixing model employed, and the last term can be omitted because advective cooling is small below the sonic point. Alternatively, one could evaluate thermal blanketing by identifying the luminosity at the photosphere \citep{Hayashi1979,Hubbard1980,Stevenson1982}. For consistency with the standard core powered mass loss model, however, we proceed with Equation~\ref{eq:luminosity} to evaluate Equation~\ref{eq:GI_EL1}.

To facilitate the comparison between Equations~\ref{eq:mass_loss_final} and \ref{eq:mass_loss_final2}, we employ the Jeans parameter,
\begin{equation}
    \Lambda = \frac{G M_{\rm n}\bar{\mu}}{k_{\rm B}T_{\rm n}R_{\rm n}},
\label{eq:Lambda}
\end{equation}
which is defined as the ratio of the gravitational potential energy to the thermal energy of a gas particle. The Jeans parameter provides a straightforward intuition for the influence of the planetary mass and surface temperature on the hydrodynamic properties of escaping winds. For atmospheric gas, $\Lambda$ has a minimum value of unity because smaller values indicate that the gas particles have sufficient thermal energy to overcome the planet's gravitational pull. Such gas is no longer gravitationally bound, so it cannot be considered part of the atmosphere. Conversely, when $\Lambda{\to}\infty$, the gas becomes increasingly gravitationally bound, making escape increasingly less likely.

To explore the relationship between the mass loss rates according to Equations~\ref{eq:mass_loss_final} and \ref{eq:mass_loss_final2} with the Jeans parameter (Equation~\ref{eq:Lambda}), we performed a numerical simulation of $2500$ synthetic exoplanets across a range of planetary parameters. The planetary masses were randomly sampled from a uniform distribution spanning $1~M_{\oplus}$ to $10~M_{\oplus}$. The atmospheric mass fractions were similarly drawn from a uniform distribution, ranging from ${-}3$ to ${-}1$ in the logarithmic space, corresponding to atmospheric masses that are $10^{-3}$ to $10^{-1}$ of the planetary mass. The atmospheric mass for each planet was then computed as the product of the planetary mass and the atmospheric mass fraction. Surface temperatures were drawn uniformly from $3000{-}10{,}000~{\rm K}$ \citep{Benz1989,Cameron1997,Canup2008,Nakajima2015,Lock2018}, and equilibrium temperatures from $500{-}3000~{\rm K}$. We removed 39 outlier planets where the radiative-convective boundary is above the sonic point or below the planetary nucleus surface.

\begin{figure}[!h]
    \centering
    \includegraphics[width=0.8\linewidth]{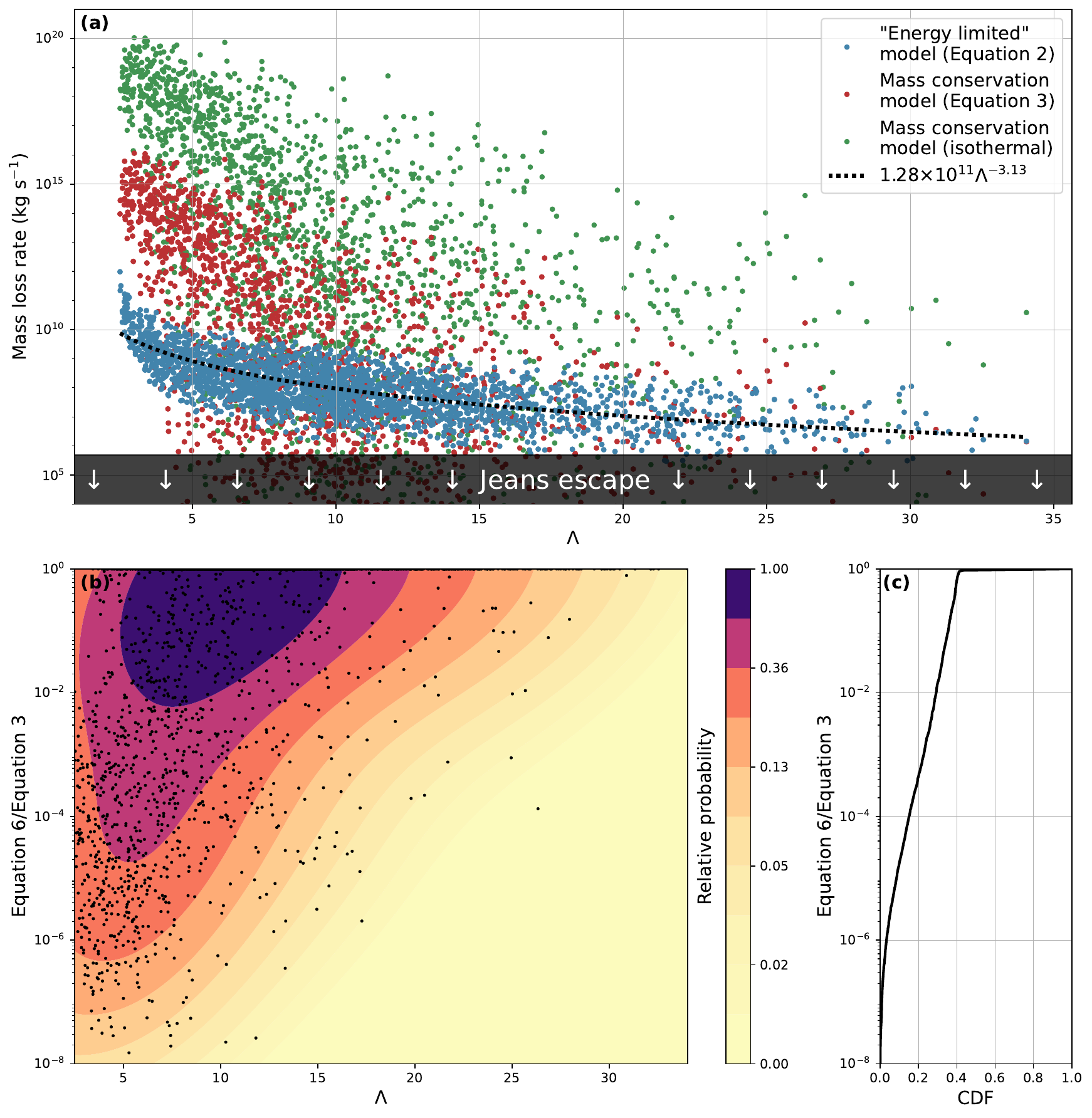}
    \caption{(a) The ``energy limited'' (Equation~\ref{eq:GI_EL1}) and the mass conservation models using Equation~\ref{eq:GI_EL2} (blue) and the isothermal hydrodynamic approximation \citep[green; e.g.,][]{Biersteker2019} as a function of the Jeans parameter at the planetary nucleus (Equation~\ref{eq:Lambda}). The ``energy limited'' model follows an exponentially decaying relation with the Jeans parameter ($R^{2}{=}0.53$ goodness fit). Mass loss rates below $5 {\times} 10^{5}~{\rm kg{\,}s^{-1}}$ are likely governed by Jeans escape rather than hydrodynamic outflow. (b) Comparison of mass loss rates calculated using the ``energy limited'' and our hydrodynamic mass conservation equation, plotted against the Jeans parameter at the planetary nucleus. We simulated $2500$ synthetic planets with masses uniformly distributed between $1{-}10$ Earth masses and atmospheric mass fractions uniformly sampled in the logarithmic space between $10^{-3}$ and $10^{-1}$. The atmospheric mass for each planet was calculated as the product of its planetary mass and atmospheric mass fraction. Surface temperatures were drawn uniformly from $3000{-}10{,}000~{\rm K}$, and equilibrium temperatures from $500{-}3000~{\rm K}$. The color shading shows relative probability density (note that only the samples with the mass loss ratio less than unity are considered here for clarity). (c) Marginalized cumulative distribution function (CDF) for the mass loss ratio of the ``energy limited'' and our hydrodynamic mass conservation equation.}
    \label{fig:ratio}
\end{figure}

Figure~\ref{fig:ratio}a shows the mass loss rates of the ``energy limited'' (Equation~\ref{eq:GI_EL1}) model, our hydrodynamic mass conservation model (Equation~\ref{eq:GI_EL2}), and the isothermal mass conservation model \citep{Biersteker2019} as a function of the Jeans parameter at the planetary nucleus (Equation~\ref{eq:Lambda}). The ``energy limited'' mass loss follows an exponentially decaying relation with the Jeans parameter whereas the mass conservation equations have a large dispersion without a clear trend. The exponential behavior of the former emerges because the ``energy limited'' equation scales with the radius of the radiative-convective boundary, which itself scales with $\Lambda/(\Lambda {-} 1)$ (see Equation~\ref{eq:T_profile}, from which $R_{\rm rcb}$ is derived). The factor $\Lambda/(\Lambda{-}1)$ diverges when $\Lambda$ approaches unity whereas the factor converges to unity as $\Lambda$ increases, creating the observed exponential-like decay. In contrast, both mass conservation mass loss models show no clear trend because their parameters are not directly related to the Jeans parameter at the planetary nucleus. The sonic point radius depends on equilibrium temperature, not magma ocean temperature, and its density is derived through integration (Equation~\ref{eq:rho_rcb}), which does not imply a simple functional relationship between density and the Jeans parameter. Missing a direct relation to $\Lambda$ results in the large dispersion in mass loss rates seen in Figure~\ref{fig:ratio}a, highlighting the inherent complexity of atmospheric loss from giant impacts. The isothermal mass conservation equation systematically overestimates the mass loss relative to our hydrodynamic model because the barometric formula erroneously converges to a finite density at infinite distance \citep[][pp.\,145$-$151]{walker1977}. However, because the standard core powered mass loss framework adopts the minimum of the isothermal mass conservation and ``energy limited'' rates, it typically defaults to the latter, which is often several orders of magnitude lower than either mass conservation estimate, significantly underestimating the atmospheric escape that can occur after a giant impact.

Figure~\ref{fig:ratio}b shows the mass loss rate ratios of Equations~\ref{eq:mass_loss_final} and \ref{eq:mass_loss_final2}, which fall into two populations: the mass loss rates are identical (i.e., min$\left[\dot{M}_{\rm GI{,}1},\dot{M}_{\rm GI{,}2}\right]/\dot{M}_{\rm GI{,}2}{=}\dot{M}_{\rm GI{,}2}/\dot{M}_{\rm GI{,}2}{=}1$) for approximately ${\sim}55\%$ of the planets (i.e., see peak in Figure~\ref{fig:ratio}c). However, for the remaining ${\sim}45\%$ (i.e., purple region in Figure~\ref{fig:ratio}b), Equation~\ref{eq:mass_loss_final} underestimates the mass loss rates by up to eight orders of magnitude compared to Equation~\ref{eq:mass_loss_final2}. The reason for this discrepancy is that the ``energy limited'' model ignores how the atmospheric density changes with radius, whereas the hydrodynamic mass conservation model explicitly accounts for density change. The density at the sonic point depends strongly on the thermal structure of the atmosphere. For example, the density of an idealized fully convecting (i.e., adiabatic) atmosphere maintains higher densities at large radii compared to an isothermal atmosphere which has an approximately exponentially decaying density profile. Because the ``energy limited'' model neglects the importance of the sonic point density, it does not capture the enhanced density and thus higher mass loss rates that occur when an atmosphere is more convecting than isothermal. This effect becomes increasingly pronounced for planets with lower Jeans parameters, which typically correspond to higher surface temperatures or lower planetary masses. Under these conditions, the radiative-convective boundary moves outward, expanding the adiabatic portion of the atmosphere, increasing the sonic point density, and substantially increasing the mass loss rates. This suggests that planets with hydrogen-rich atmospheres can lose their atmospheres much more rapidly after a giant impact than previously assumed. Contrary to the standard framework suggesting that total mass loss occurs over million year timescales \citep{Ginzburg2018,Gupta2018,Biersteker2019,Gupta2020}, the revised framework implies that mass loss can occur almost instantaneously after a giant impact \citep[i.e., regimes one and two of][]{Modirrousta2023}.

\section{Developing an efficient approximation}
\label{sec:approximation}

Our revised hydrodynamic framework (Equation~\ref{eq:GI_EL2}) improves significantly on both the ``energy limited'' and isothermal mass conservation models (Equation~\ref{eq:mass_loss_final}). The ``energy limited'' model incorrectly assumes that escape is governed by internal luminosity at the radiative-convective boundary, whereas the isothermal model adopts an unphysical density structure that does not account for adiabatic cooling and approaches a finite value at infinite radius. Our model incorporates the thermal structure of the outflow and provides a more realistic description of mass loss following a giant impact. Our approach is well suited for individual planets or small samples, where accuracy is important. However, it is computationally demanding for large populations. Computing the mass loss rate requires solving coupled equations for velocity (Equation~\ref{eq:v_sol}), density (Equation~\ref{eq:rho_profile}), and temperature (Equation~\ref{eq:T_profile}), as well as integrating to find the radiative-convective boundary density (Equation~\ref{eq:rho_rcb}). For population studies involving tens to hundreds of thousands of planets, solving this full set of equations for each planet can be computationally demanding.

We therefore introduce a physically motivated approximation to Equation~\ref{eq:GI_EL2} that maintains accuracy while being significantly more efficient. We start by expressing Equation~\ref{eq:GI_EL2} as,
\begin{equation}
    \frac{{\rm d}M_{\rm atm}}{{\rm d}t} = C\left(\frac{R_{\rm s}}{R_{\oplus}}\right)^{2} \rho_{\rm s}^{\ast}(R_{\rm s}),
\label{eq:non_dim}
\end{equation}
where $C$ is a collection of constants and $\rho_{\rm s}^{\ast}(R_{\rm s})$ is the nondimensionalized sonic point density, which must depend on $R_{\rm s}$ because it defines the volume over which the atmospheric mass is distributed. To quantify this relationship and find the value of $C$, we simulate atmospheric escape for 2500 synthetic exoplanets using the same setup and input conditions described in Section~\ref{sec:outlook}. For each planet, we solve the full hydrodynamic equations to determine the sonic point radius and corresponding density, and then compute the resulting mass loss rate from Equation~\ref{eq:GI_EL2} to derive a best-fit form for Equation~\ref{eq:non_dim}.
\begin{figure}[!h]
    \centering
    \includegraphics[width=0.7\linewidth]{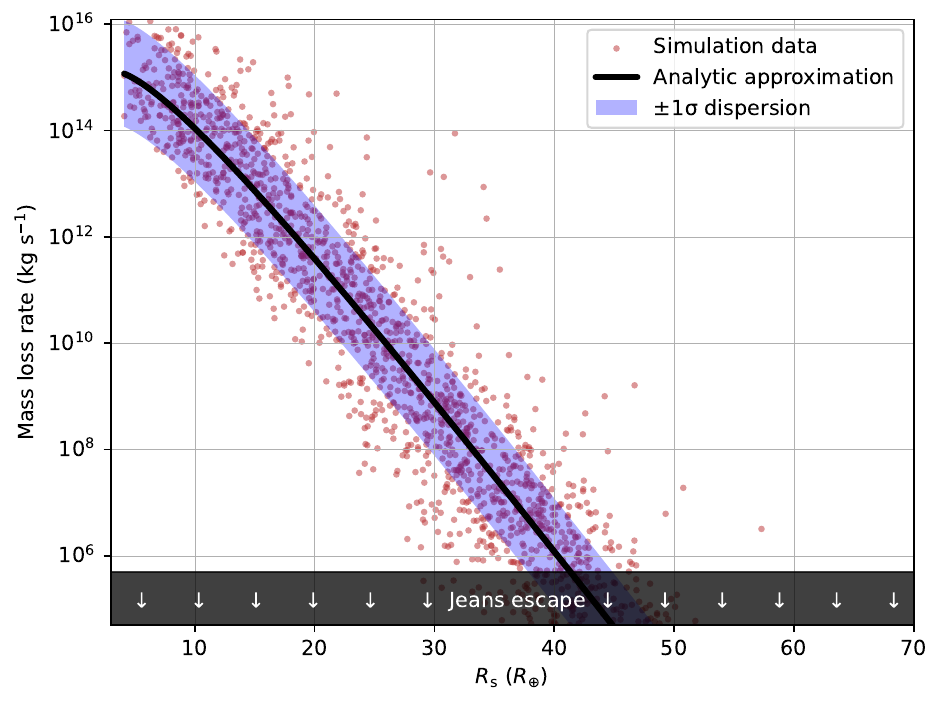}
    \caption{The mass loss rate vs the sonic point radius of 2500 synthetic planets. The red points are the mass loss rates from our hydrodynamic simulations, the black line is the analytic fit (Equation~\ref{eq:fit}), and the blue shaded region is the one standard deviation scatter in the simulated mass loss rates relative to the fit. Mass loss rates below $5 {\times} 10^{5}~{\rm kg{\,}s^{-1}}$ are likely governed by Jeans escape rather than hydrodynamic outflow.}
    \label{fig:fit}
\end{figure}

Figure~\ref{fig:fit} shows the mass loss rate for each planet plotted against their corresponding sonic point radius. We find that the mass loss rate is well approximated by (goodness fit $R^{2}{=}0.93$),
\begin{equation}
    \frac{{\rm d}M_{\rm atm}}{{\rm d}t}=1.26 \times 10^{15{\pm}1}\left(\frac{R_{\rm s}}{R_{\oplus}}\right)^{2}\exp\left({-}\frac{19}{27}\frac{R_{\rm s}}{R_{\oplus}}\right) \, \mathrm{{kg/s}},
\label{eq:fit}
\end{equation}
where the ${\pm}1$ accounts for the one standard deviation dispersion in the simulated mass loss rates relative to Equation~\ref{eq:fit}. Whereas our fit provides a more physically motivated alternative to the ``energy limited'' model, it can still deviate from the true mass loss rate by one order of magnitude or more. Therefore, it should be used with caution and not as a replacement for the full hydrodynamic treatment when precision is needed.

\section{Conclusions}
\label{sec:conclusions}

In this paper, we have shown that core powered mass loss is not limited by the internal luminosity traversing through the radiative-convective boundary. First, the magma ocean does not cool efficiently when atmospheric loss occurs. Second, for sufficiently high surface temperatures following a giant impact, the planetary atmosphere has a temperature profile that is inherently unstable. In this state, atmospheric loss proceeds through the conversion of potential energy stored in the atmosphere into work done against gravity as gas is transported to the sonic point. The outflow remains in steady state because radiative diffusion transfers energy efficiently within the atmosphere and does not impose an energy bottleneck. Third, we showed that whereas the isothermal mass conservation equation can overestimate the escape rate by neglecting adiabatic cooling, the standard core powered mass loss framework adopts the minimum of the ``energy limited'' and isothermal mass conservation rates, and thus typically defaults to the former. As a result, the standard core powered model may underestimate the atmospheric escape after a giant impact by up to eight orders of magnitude. To address this, we developed a full hydrodynamic model that captures the effects of adiabatic cooling and internal atmospheric structure, overcoming the limitations of both existing approaches. Moreover, we provide a physically motivated analytic fit to our hydrodynamic solution, intended for evaluating the mass loss rate of exoplanets in large-scale population studies. In short, by eliminating the ``energy limited'' argument and correcting the isothermal assumption, we offer a revised framework that simplifies the understanding of atmospheric escape following giant impacts. These findings have broad implications for interpreting planetary evolution and atmospheric retention.

\section*{Declaration of Competing Interest}
The authors have no competing interests.

\section*{Data availability}
No data was generated in the making of this paper.

\section*{Acknowledgements}
This work was sponsored by the US National Science Foundation EAR-2224727 and the US National Aeronautics and Space Administration under Cooperative Agreement {No.\,80NSSC19M0069} issued through the Science Mission Directorate. We thank the two anonymous referees for their constructive comments, which informed us of relevant literature and helped us refine our hydrodynamic framework.

\appendix
\renewcommand{\theequation}{A\arabic{equation}}
\setcounter{equation}{0}  
\renewcommand{\thesection}{A\arabic{section}}
\setcounter{section}{0}  

\section{Derivation of the temperature gradient and profile}
\label{sec:dT_dr}

Solving the ideal gas equation for temperature, $T{=}P\bar{\mu}/\left(\rho k_{\rm B}\right)$, and differentiating with respect to radial distance yields,
\begin{equation}
    \frac{{\rm d}T}{{\rm d}r} = \frac{\bar{\mu}}{\rho k_{\rm B}}\frac{{\rm d}P}{{\rm d}r} - \frac{P \bar{\mu}}{\rho^{2}k_{\rm B}}\frac{{\rm d}\rho}{{\rm d}r},
\end{equation}
where $P$ is pressure, $\rho$ is density, $T$ is temperature, $\bar{\mu}$ is the mean molecular mass, and $k_{\rm B}$ is the Boltzmann constant. Differentiating the isentropic gas relation, $P{\propto}\rho^{\gamma}$, where $\gamma$ is the isentropic exponent (i.e., heat capacity ratio), with respect to $r$, one obtains,
\begin{equation}
    \frac{{\rm d}P}{{\rm d}r} = \frac{\gamma P}{\rho}\frac{{\rm d}\rho}{{\rm d}r},
\end{equation}
which can be combined with the temperature gradient to arrive at,
\begin{equation}
    \frac{{\rm d}T}{{\rm d}r} = \frac{\gamma-1}{\gamma}\frac{\bar{\mu}}{\rho k_{\rm B}}\frac{{\rm d}P}{{\rm d}r}.
\label{eq:dT_dr_app}
\end{equation}
Next, we consider momentum conservation for a steady-state, spherically symmetric radial outflow. In spherical coordinates, this can be written as,
\begin{equation}
\frac{{\rm d}P}{{\rm d}r} = \left.\frac{{\rm d}P}{{\rm d}r}\right|_{\rm hs} + \left.\frac{{\rm d}P}{{\rm d}r}\right|_{\rm hd},
\end{equation}
where $\left.{\rm d}P/{\rm d}r\right|_{\rm hs} {=} {-}\rho g$ is the hydrostatic term, and $\left.{\rm d}P/{\rm d}r\right|_{\rm hd} {=} {-}\rho v \left({\rm d}v/{\rm d}r\right)$ is the hydrodynamic (advective) term. Here, $g$ is the gravitational acceleration and $v$ is the radial advection velocity. Following the core powered mass loss framework, we assume that the upper atmosphere is nearly isothermal, with deviations from isothermality arising only from adiabatic expansion associated with the accelerating outflow. In other words, convective and other mixing processes (e.g., eddy diffusion) do not contribute significantly to cooling. Thus, only the hydrodynamic component of the pressure gradient enters the temperature gradient. Substituting $\left.{\rm d}P/{\rm d}r\right|_{\rm hd}$ into Equation~\ref{eq:dT_dr_app} yields,
\begin{equation}
    \frac{{\rm d}T}{{\rm d}r} = - \frac{\gamma - 1}{\gamma} \frac{\bar{\mu} v}{k_{\rm B}} \frac{{\rm d}v}{{\rm d}r}.
\end{equation}

To derive Equation~\ref{eq:T_ratio}, we integrate the temperature gradient between the radiative-convective boundary $(T_{\rm rcb}, v_{\rm rcb})$ and the sonic point $(T_{\rm s}, c_{\rm s})$,
\begin{equation}
    T_{\rm s} - T_{\rm rcb} = - \frac{\gamma - 1}{2} \frac{\bar{\mu}}{\gamma k_{\rm B}} \left(c_{\rm s}^{2}-v_{\rm rcb}^{2}\right).
\end{equation}
Dividing both sides by $T_{\rm s}$ and substituting $c_{\rm s} {\equiv} \sqrt{\gamma k_{\rm B} T_{\rm s} / \bar{\mu}}$, we obtain,
\begin{equation}
    1 - \frac{T_{\rm rcb}}{T_{\rm s}} = - \frac{\gamma - 1}{2} \left[1-\left(\frac{v_{\rm rcb}}{c_{\rm s}}\right)^{2}\right].
\end{equation}
Rearranging terms,
\begin{equation}
    \frac{T_{\rm rcb}}{T_{\rm s}} = \frac{\gamma + 1}{2} - \frac{\gamma-1}{2}\left(\frac{v_{\rm rcb}}{c_{\rm s}}\right)^{2},
\end{equation}
and inverting, yields our final functional form,
\begin{equation}
    \frac{T_{\rm s}}{T_{\rm rcb}} = \frac{2}{\gamma + 1 - \left(\gamma - 1\right)\left(\frac{v_{\rm rcb}}{c_{\rm s}}\right)^{2}}.
\end{equation}

\section{Derivation and uniqueness of the temperature profile}
\label{sec:unique}

In our mass loss model, the temperature profile must satisfy the boundary conditions $T(0)/T_{\rm s}{=}\left(\gamma{+}1\right)/2$, $T(R_{\rm s})/T_{\rm s}{=}1$, and $T(r{\to}\infty)/T_{\rm s}{\to}0$. For the latter condition to hold, the right-hand side of Equation~\ref{eq:revealing_equation_3} must tend to zero, requiring $T(r) {\propto} 1/r$ for large r, and therefore ${\rm d}T/{\rm d}r {\propto} {-}T^2$ to leading order. Many functions share this asymptotic form, so it does not uniquely determine the behavior of the temperature at all radii. To construct a tractable analytic model, we adopt the differential equation,
\begin{equation}
    \frac{{\rm d}T}{{\rm d}r} = -\beta T^2,
\label{eq:ODE}
\end{equation}
with $\beta$ a constant to be determined. Solving Equation~\ref{eq:ODE} with the boundary conditions of $T(0)/T_{\rm s}{=}\left(\gamma{+}1\right)/2$, $T(R_{\rm s})/T_{\rm s}{=}1$, and $T(r{\to}\infty)/T_{\rm s}{\to}0$, we obtain,
\begin{equation}
    \frac{T(r)}{T_{\rm s}}=\frac{1}{1+\beta T_{\rm s}(r-R_{\rm s})},
\end{equation}
where,
\begin{equation}
    \beta=\frac{\gamma-1}{\left(\gamma+1\right)R_{\rm s}T_{\rm s}}.
\end{equation} 
To confirm that this solution is unique for the differential equation above, we apply the Picard–Lindelöf theorem for the initial value problem of Equation~\ref{eq:ODE}. The theorem guarantees a unique solution provided the right-hand side function, $f(T){=}{-}\beta T^2$, is continuous and Lipschitz continuous in $T$. A function is Lipschitz continuous on a given interval if there exists a constant $\mathcal{L}$ (i.e., the Lipschitz constant) such that for any two values $T_{1}$ and $T_{2}$ within that interval, $|f(T_{1}){-}f(T_{2})| {\leq} \mathcal{L}|T_{1}-T_{2}|$. This condition ensures that a small change in $T$ produces a proportionally small change in $f(T)$, and the function does not change too abruptly. Because $f(T)$ is a polynomial with respect to $T$, it is continuous everywhere. Next, to show Lipschitz continuity, we consider two arbitrary temperatures $T_{1}$ and $T_{2}$ within the physically relevant range $0{\leq} T {\leq} \left(\gamma{+}1\right)T_{\rm s}/2$. Then, $|f(T_1){-}f(T_2)|{=}\beta|T_1^{2}{-}T_2^{2}|{=}\beta|T_{1}{+}T_{2}||T{_1}{-}T{_2}|$. Because $T_{1}$ and $T_{2}$ are bounded from above by $\left(\gamma{+}1\right)T_{\rm s}/2$, we have $|T_{1}{+}T_{2}|{\leq}\left(\gamma{+}1\right)T_{\rm s}$. Thus, $|f(T_{1}){-}f(T_{2})|{\leq} \beta \left(\gamma{+}1\right)T_{\rm s}|T_{1}{-}T_{2}|$,
which shows that $f(T)$ is Lipschitz continuous with Lipschitz constant $\mathcal{L}{=}\beta\left(\gamma{+}1\right)T_{\rm s}$. The Picard–Lindelöf theorem then guarantees that this solution is unique for the initial condition at $r{=}R_{\rm s}$ and for the assumed global form of the temperature gradient (Equation~\ref{eq:ODE}).

To evaluate how temperature affects the last term of Equation~\ref{eq:revealing_equation_3}, we compare three temperature profiles: (1) our analytic solution (Equation~\ref{eq:T_solution}), (2) a semi-isothermal approximation, $T(r){=}T_{\rm s}$ (applied only to the rightmost term of Equation~\ref{eq:revealing_equation_3}), and (3) a linear approximation, $T(r)/T_{\rm s}{=}(\gamma{+}1)/2{-}(\gamma{-}1)(r/R_{\rm s})/2$. Our analytic profile satisfies all three boundary conditions; the linear approximation satisfies two, $T(0)/T_{\rm s}{=}(\gamma{+}1)/2$ and $T(R_{\rm s})/T_{\rm s}{=}1$; and the semi-isothermal approximation satisfies only one, $T(R_{\rm s})/T_{\rm s}{=}1$.
\begin{figure}[!h]
\centering
\includegraphics[width=0.6\linewidth]{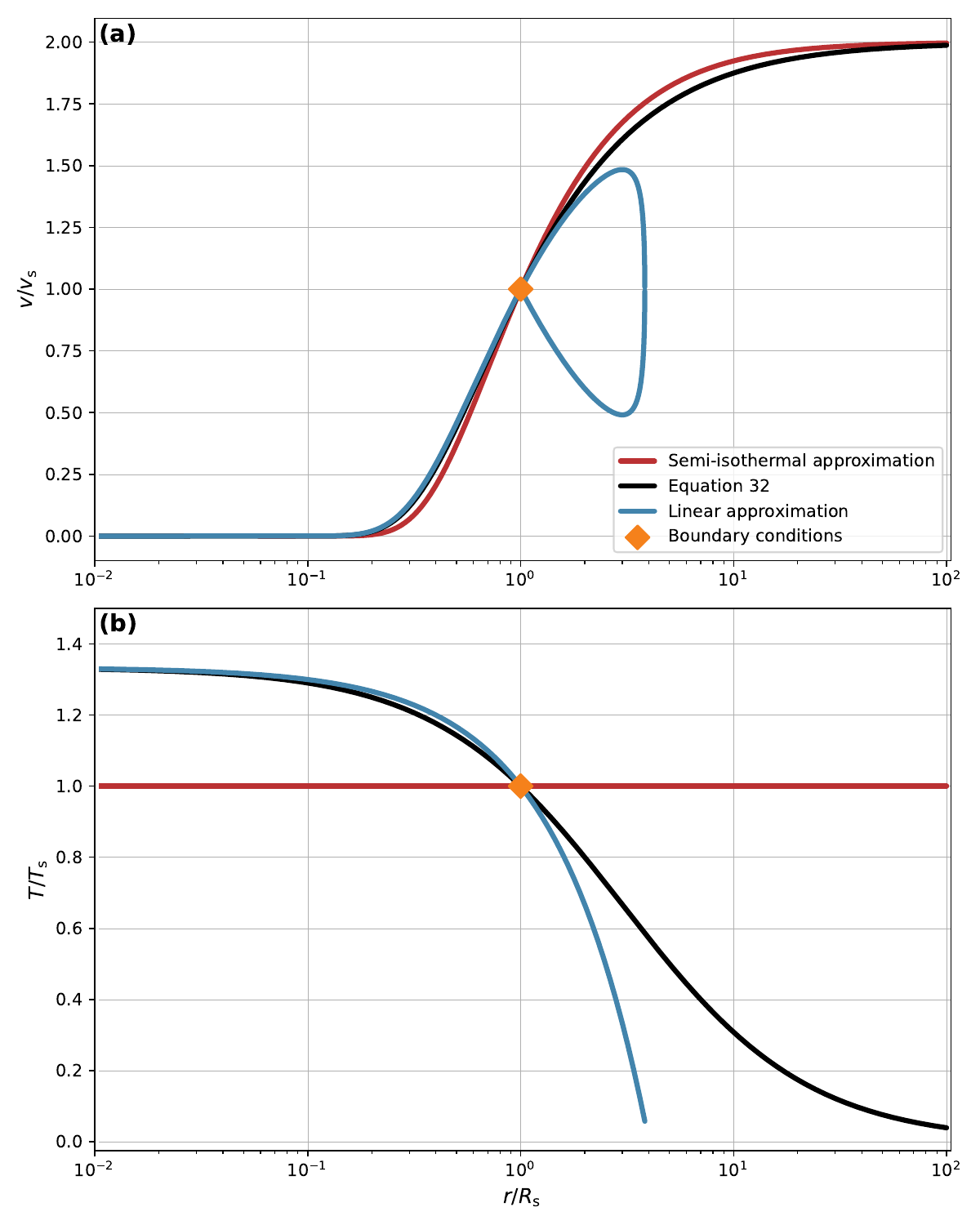}
\caption{(a) Velocity profiles for three temperature models: analytic solution (Equation~\ref{eq:T_solution}; black), semi-isothermal approximation $T(r){=}T_{\rm s}$ (red), and linear approximation $T(r)/T_{\rm s}{=}(\gamma{+}1)/2 {-} (\gamma{-}1)(r/R_{\rm s})/2$ (blue) for $\gamma{=}5/3$. Orange diamonds indicate boundary condition at $r{=}R_{\rm s}$ and $v{=}v_{\rm s}$. The boundary conditions at $r{=}0$ ($v{=}0$) and $r{\rightarrow}\infty$ ($v{\rightarrow}\sqrt{(\gamma+1)/(\gamma-1)}v_{\rm s}$) are not shown. (b) Corresponding temperature profiles for the three models.}
\label{fig:benchmark}
\end{figure}
Figure~\ref{fig:benchmark} shows the three temperature profiles and their corresponding velocity profiles for $\gamma{=}5/3$. The linear approximation does not yield valid solutions across all radii and it is therefore unphysical. Our analytic solution and the semi-isothermal approximation perform better, with the semi-isothermal model producing lower velocities for $r/R_{\rm s}{<}1$ and higher velocities for $r/R_{\rm s}{>}1$, corresponding to when $T(r){>}T_{\rm s}$ and $T(r){<}T_{\rm s}$, respectively. Whereas the semi-isothermal model offers a simple approximation that adequately captures the expected velocity profile, only our analytic solution satisfies all boundary conditions and yields a unique, globally consistent temperature profile. This justifies the use of Equation~\ref{eq:T_solution} in the main text.

\printcredits

\bibliographystyle{cas-model2-names}

\bibliography{cas-refs}



\end{document}